# Electrocatalytic reduction of nitrogen to ammonia in ionic liquids

Yuan Tian,[a,b] Yanrong Liu,[b,c] Hao Wang,[b] Lei Liu,[b,d*] and Wenping Hu[a,*]

[a] Tianjin Key Laboratory of Molecular Optoelectronic Sciences, Department of Chemistry, School of Science, Tianjin University, Tianjin 300072, China
[b] Beijing Key Laboratory of Ionic Liquids Clean Process, CAS Key Laboratory of Green Process and Engineering, State Key Laboratory of Multiphase Complex Systems, Institute of Process Engineering, Chinese Academy of Sciences, Beijing 100190, China
[c] Zhengzhou Institute of Emerging Industrial Technology, Zhengzhou 45000, China
[d] School of Chemistry and Chemical Engineering, Wuhan Textile University, Wuhan, 430200, China

Corresponding authors:
Lei Liu, 2021047@wtu.edu.cn; liulei3039@gmail.com
Wenping Hu, huwp@tju.edu.cn

**Abstract:** Ammonia ($NH_3$) is an important raw material for nitrogen fertilizer production and is widely used in industry. It is also a carbon-free renewable fuel and an excellent hydrogen storage material. Reaction conditions of traditional Haber–Bosch process are very harsh although it supports most of today's $NH_3$ production. Recently, electrocatalytic nitrogen reduction reaction (NRR) has become one of the effective alternative methods to the traditional Haber–Bosch process due to the advantages of mild operation conditions and almost no environmental pollutions. Ionic liquids (ILs) act as one of green solvents possess a lot of unique features and have shown excellent performance in electrocatalytic NRR process. In this review, we summarize fundamentals of the electrocatalytic NRR, the role of ILs in several typical electrocatalytic processes, and the possible role of ILs in the electrocatalytic NRR process. Moreover, the selection principles and issues of the application of ILs, as well as the future development direction for the electrocatalytic NRR process are concluded.

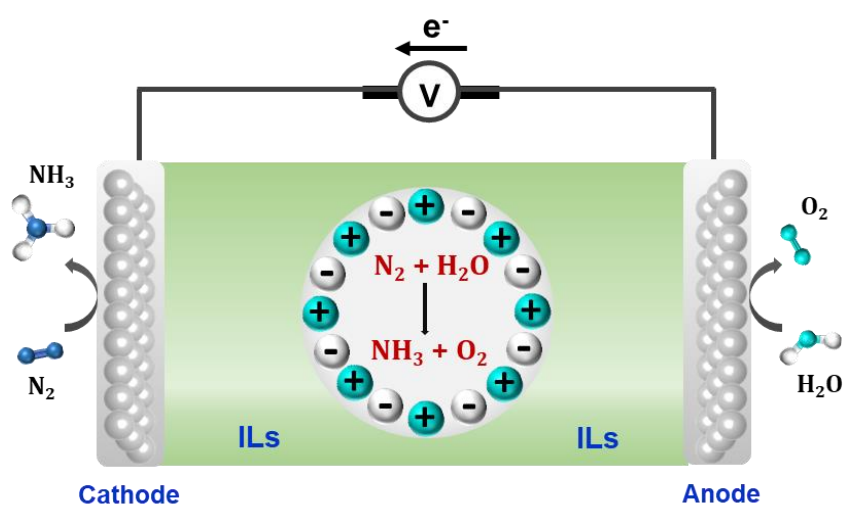

**TOC:** This review summarizes fundamentals of the electrocatalytic NRR, and the role of ILs in the electrocatalytic NRR process.

# 1. Introduction

Ammonia ($NH_3$) is a kind of important raw material for nitrogen fertilizer production, as well as is widespread applied in the industry, including explosives, fuels, fibers, resins and so on.[1-4] As one of the most important substances, $NH_3$ has played an important role in promoting the revolution of the industry and agriculture, and has pivotal functions for national economies.[5] Moreover, $NH_3$ is a carbon-free renewable fuel and is an excellent hydrogen storage material, hence, it attracts a lot of attention from researchers.[6, 7] $NH_3$ synthesis is mainly realized by $N_2$ fixation, and the $N_2$ fixation was almost all realized by biological nitrogen fixation before the discovery of Haber–Bosch process.[8, 9] In spite of there are 78 vol% $N_2$ in the air, but it is very difficult to fix nitrogen from the atmosphere due to the inertia of nitrogen molecules. Scientists never stopped the steps to explore effective methods of synthesizing $NH_3$, *e.g.*, German chemist Fritz Haber started to investigate the $NH_3$ synthesis from 1902, and achieved a high yield of $NH_3$ by optimizing the reaction conditions in 1909. That is the famous Haber–Bosch process, the first economically feasible method of directly synthesizing $NH_3$ from hydrogen and atmospheric nitrogen.[10] However, the reaction conditions of traditional Haber–Bosch process are very harsh with the temperature of 300~500 °C and the pressure of 100~200 atm, causing a lot of energy loss which is about 2% of the world's total energy consumption.[4, 10, 11] Meanwhile, the clean $H_2$ is often obtained from fossil and result in a lot of $CO_2$ emissions ($C+2H_2O \rightarrow 2H_2+CO_2$).[12-15] Therefore, great efforts have been made to explore alternative processes of $NH_3$ synthesis under mild conditions.

Currently, the methods of preparing $NH_3$ artificially under mild conditions mainly include biological enzymes $N_2$ fixation, photocatalytic, and electrocatalytic $NH_3$ synthesis. In the 1960s, enzymes FeMo was used to fix $N_2$ under mild conditions. The biological azotase can reduce $N_2$ to $NH_3$ at room temperature and under normal pressure, and without the consumption of $H_2$. However, the biological $N_2$ fixation process is slow, and is sensitive to the environment.[4] The photocatalytic $N_2$ reduction relies on solar energy and the unique properties of semiconductor materials, however, the utilization of proton is low which is owing to rapid recombination of photogenerated carriers.[16, 17] The electrocatalytic $N_2$ reduction is driven by electrons, takes $H_2O$ and $N_2$ as raw materials, and becomes one of the effective alternative processes to the traditional

Haber–Bosch process for $NH_3$ synthesis, which can break the thermodynamic equilibrium and realize the thermodynamic nonspontaneous $NH_3$ synthesis reactions by the action of electric energy.[18, 19] Electrocatalytic nitrogen fixation could be date back to the year of 1969, when Van et al.[20] from Stanford University studied the nitrogen fixation performance in the Ti-Al system. The reaction of electrocatalytic NRR to synthesis $NH_3$ under ambient conditions is often restricted by the low $NH_3$ yields and low Faradaic Efficiency (FE). On the one hand, nitrogen is an inert gas and the triple bond is very stable, thus, the cleavages and activated of nitrogen-nitrogen triple bond is difficult.[21, 22] On the other hand, the solubility of $N_2$ in water is very low, and the competition reaction of hydrogen evolution reaction (HER) could react under lower potential compared with NRR, hence, the generation of $H_2$ is easier than $N_2$.[23-25]

At present, the main strategies to improve the reaction activity and FE include two aspects: (1) NRR catalysts design which mainly include the structure optimization,[26-29] heteroatomic doping,[25, 30] space engineering,[31] surface/interface engineering[32] and synergistic effects;[33] (2) Catalytic system integration, mainly include the modification of the electrolytes. Often the specific ions are introduced into the electrolyte or add ionic liquids (ILs) to electrolytes, even the pure ILs are used as electrolytes to improve the solubility of $N_2$ or suppress competition reaction of HER,[34-36] as well as the improvement of $NH_3$ yields. There are several review papers summarizing the NRR process,[1, 37-41] however, the combination of electrocatalytic NRR process and ILs are somehow few. In this review, we focus on the electrocatalytic NRR and the role of ILs in the process of electrocatalytic NRR. Firstly, we discuss the basic information for the heterogeneous electrocatalytic process of NRR, including the reaction process, competition reactions, the recognized reaction pathways as well as the development of theoretical calculations. Secondly, the roles of ILs in electrocatalytic process including $CO_2$ reduction reaction ($CO_2RR$), oxygen reduction reaction (ORR) are summarized. Then the application and possible mechanism of ILs in the electrocatalytic NRR are discussed. Lastly, we conclude the selection principles and the issues of the application of ILs, as well as the future development direction for the electrocatalytic NRR process.

## 2. Electrocatalytic NRR in aqueous solution

### 2.1 Elementary processes

$N_2$ is a highly stable molecule with a strong triple bond being 941 kJ mol$^{-1}$ and a large HOMO-LUMO gap being 10.82 eV.[40] Hence, the nitrogen-nitrogen triple bond is difficult to be broken. The heterogeneous electrocatalytic process for NRR is based on the electron and proton charge transfer, where external voltage is applied to adjust the reaction activation energy of electrode surface, and to speed up the reaction process under ambient conditions. Often, a three-electrode test system is used for electrochemical NRR process, in which the anode and cathode electrodes are separated by a proton exchange membrane or anion exchange membrane. According to the experimentation procedure, the electrolytic cell can be divided to four types, including back-to-back cell, PEM-type cell, H-type cell and single chamber cell (See Fig. 1)[1]

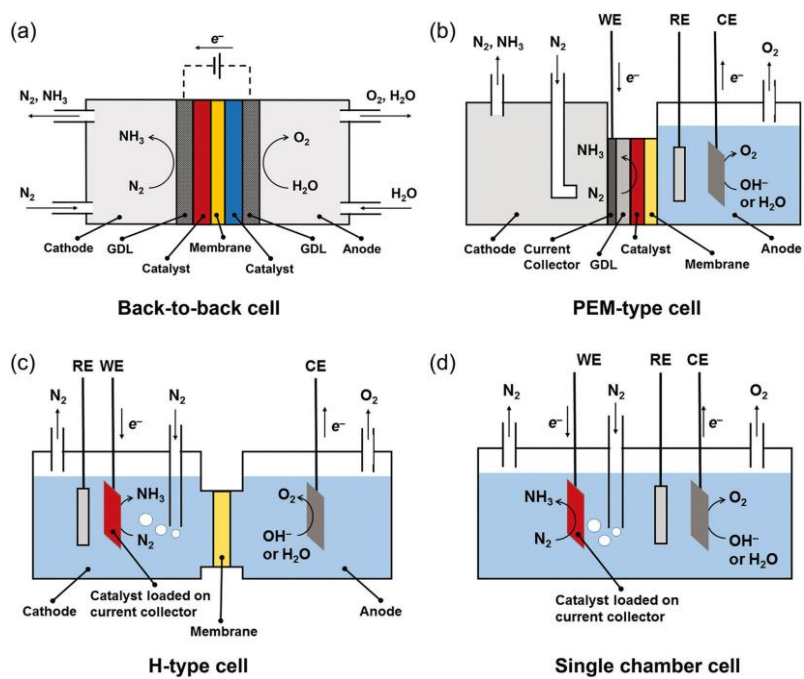

**Fig. 1** Schematic diagrams of different cell types for electrocatalytic NRR at ambient conditions.[1] Reproduced with permission from ref. 1. Copyright 2018 Wiley.

In all used electrolytic cells, the anodic reaction is the $O_2$ producing reaction, while the cathodic reaction is the reduction of $N_2$ (see Equations 1~7). As shown in Equations 2, 3 and 5, 6, the standard electrode potential of $H_2$ generation is similar to $NH_3$ generation under both acidic and alkaline conditions, making HER become the main competing reaction. Moreover, it can be found that the HER is thermodynamically similar to the NRR, even less negative HER limiting-potential requirements on many potentially active metal catalysts than that of NRR.[42] It is unavoidable for the generation of $H_2$ during electrocatalytic NRR process, and this is also a bottleneck for the development of electrocatalytic NRR to synthesis $NH_3$.

Acidic condition (pH = 0):

    Anodic reaction      $3H_2O \rightarrow 3/2O_2\,(g)+6H^++6e^-$, $E^0$=1.23 V *vs.* RHE      [1]

    Cathodic reaction      $N_2\,(g)+6H^++6e^- \rightarrow 2NH_3\,(g)$, $E^0$=-0.148 V *vs.* RHE      [2]

    Competition reaction      $2H^++2e^- \rightarrow H_2\,(g)$, $E^0$=0 V *vs.* RHE      [3]

Alkaline condition (pH = 14):

    Anodic reaction      $6OH^- \rightarrow 3H_2O+3/2O_2\,(g)+6e^-$, $E^0$=0.401 V *vs.* RHE      [4]

    Cathodic reaction    $N_2\,(g)+6H_2O+6e^- \rightarrow 2NH_3\,(g)+6OH^-$,

                                                        $E^0$=-0.736 V *vs.* RHE      [5]

    Competition reaction      $2H_2O+2e^- \rightarrow H_2\,(g)+2OH^-$, $E^0$=-0.828 V *vs.* RHE      [6]

Overall reaction:      $N_2\,(g)+3H_2O \rightarrow 2NH_3\,(g)+3/2O_2\,(g)$      [7]

## 2.2 Reaction pathways

As a typical heterogeneous catalytic process, the reaction mechanism of Haber–Bosch process has been studied extensively. Ertl et al.[43, 44] used scanning tunneling microscopy and electron spectroscopic methods to obtain direct atomic resolution images for studying detailed information about the influence of catalyst and the bonding state of adsorbed species. As shown in Fig. 2, the authors compared the $NH_3$ synthesis process with and without Fe catalyst, and concluded that the $N_2$ and $H_2$ are firstly dissolved into chemisorbed *N and *H, and then the reaction between *N and *H would be continuously taken until the formation and desorption of $NH_3$. The presence of Fe catalyst greatly promoted the reaction process.

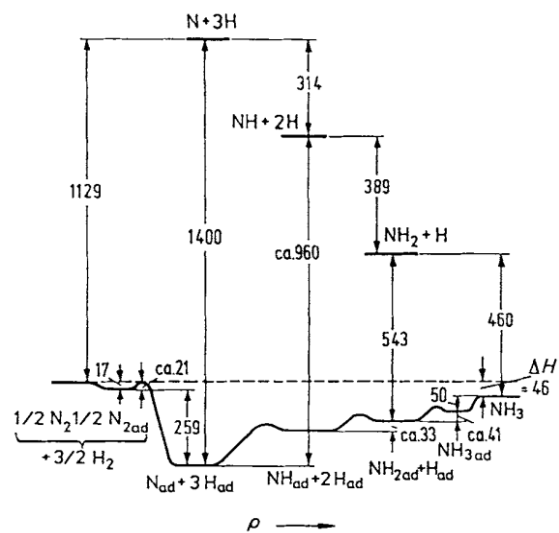

**Fig. 2** Schematic potential diagram of the energy differences for the NH$_3$ synthesis by Fe-based catalyst in comparison to the non-catalytic steps (energy in kJ mol$^{-1}$).[43] Reproduced with permission from ref. 43. Copyright 1990 Wiley.

Currently, the commonly accepted reaction pathways of electrocatalytic NRR include dissociative pathway and associative pathway, and the associative pathway is future divided into a distal pathway and an alternating pathway. The dissociative pathway is shown in Fig. 3a, the first step is the broken of nitrogen-nitrogen triple bond to form dissociative *N, and the two dissociative *N are adsorbed onto the surface of metal catalyst, and then each hydrogenation process takes place independently. Unlike the dissociative pathway, the associative pathway is that the N$_2$ will be adsorbed directly on the surface of catalyst rather than be directly cleaved for the first step. During the whole hydrogenation processes, the bonds between nitrogen-nitrogen become weaker and weaker, and the completely broken of bonds between nitrogen-nitrogen occurs until the first NH$_3$ formation. As shown in Fig. 3b, the distal pathway is that the first three protons will combine with the end N absolutely until the first NH$_3$ formation and desorbed from the surface of metal catalyst, the remaining *N will proceed the follow-up hydrogenation processes. As shown in Fig. 3c, the alternating pathway is that the protons will combine with the two N in turn and the formation of the two NH$_3$ almost at the same time. It is worth mentioning that the simultaneous adoption of two N on the surface of metal catalysts is called enzymatic catalysis mechanism[45] which also can be recognized as particular associative pathway (as shown in Fig. 3bII and 3cII).

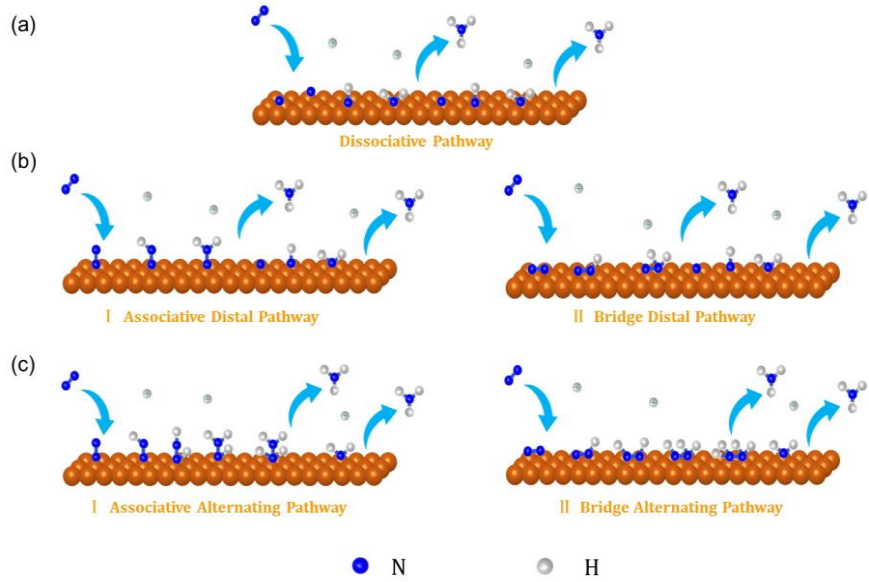

**Fig. 3** The schematic diagrams of main reaction pathways of electrocatalytic NRR at ambient conditions.

Recently, a new pathway named Mars–van Krevelen (MvK) pathway was proposed by Abghoui and Skúlason.[46] In that work, the authors screened for a stable and active catalyst amongst a range of rocksalt [RS(111)] transition metal nitrides (TMN) surfaces with NaCl-type structure, and studied the catalyzing electrochemical nitrogen activation and $NH_3$ synthesis through theoretical calculations. They proposed a heterogeneous MvK mechanism, of which two surface N atoms are reduced to generate two $NH_3$ and the remained nitrogen vacancies (dimer N-vacancy) were then filled with solvated $N_2$ in the electrolyte, then the hydrogen atoms are added to the surface one by one (as shown in Equation 8 and Fig. 4).

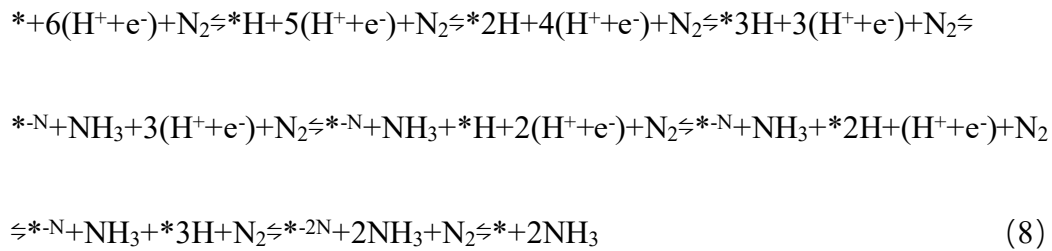

(8)

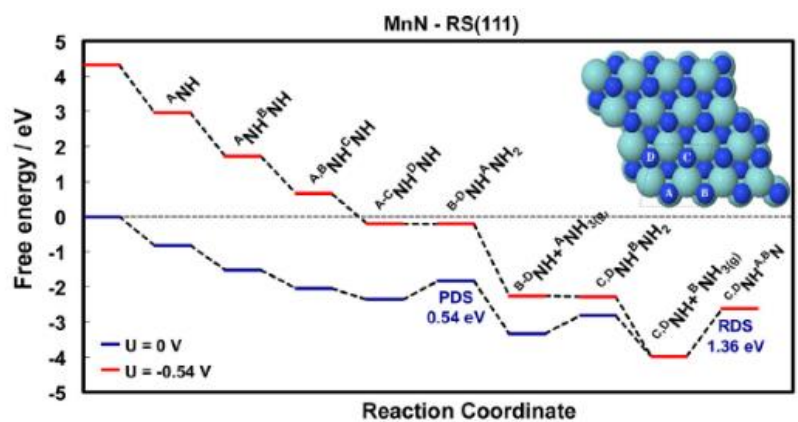

**Fig. 4** The heterogeneous MvK mechanism of electrocatalytic NRR on MnN.[46] Reproduced with permission from ref. 46. Copyright 2017 Elsevier.

**2.3 Theoretical studies**

For the complex electrocatalytic processes, one of the key problems is to seek for highly efficient and stable catalysts. Over the past few decades, many researches focused on the electrocatalytic mechanism, however, most of the researches still lack of in-depth and clear understanding, including the catalysts surface active sites and the whole reaction processes. The aim of theoretical calculations is to establish a composition-structure-function relationship in order to effectively guide experimental investigation.[47] In 2012, Nørskov et al.[48] developed the famous synthesis $NH_3$ volcano diagrams through density functional theory (DFT). They revealed the relationship between surface adsorption energy and catalytic performance in acid electrolyte. DFT takes descriptors to calculate the bond energy and strength quantitatively, and obtain various transient states results. All the intermediates are achieved by *N combining with catalyst surface, and the related adsorption energies are linear dependence with binding energies of *N (as shown in Fig. 5a-d). The suitable catalysts could be screened by comparing the energy barriers, as well as the reaction processes will be predicted and the volcano diagrams will be built. As shown in Fig. 5e, the metals located at different locations are with different catalysis performances and different selectivity of NRR and HER. The right legs of the volcanoes have the rate-determining of $N_2$ splitting and the activity of the metals is determined by the adsorption of $N_2$ and the first proton transfer step (*$N_2H$). The metals on the left legs have the rate-determining steps of *NH to *$NH_2$ and *$NH_2$ to $NH_3$.

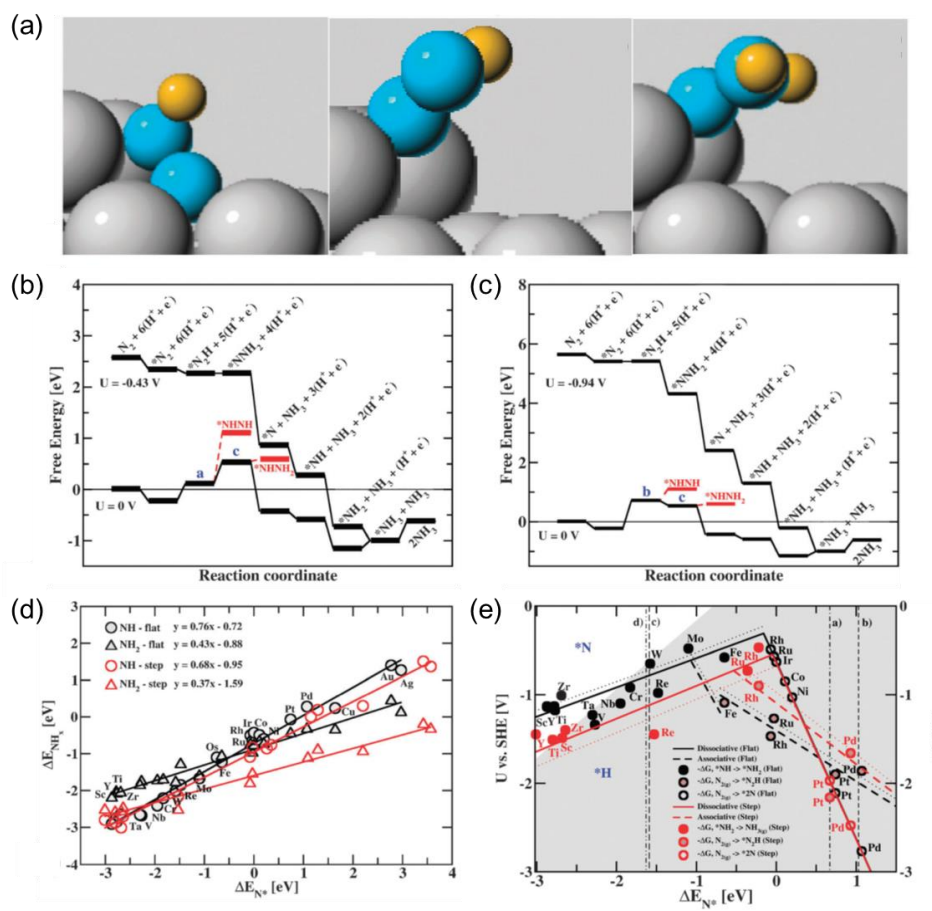

**Fig. 5** (a) Minimum energy configurations of *$N_2$H and *$NNH_2$; (b, c) Free energy for associative mechanism at the step on the Ru(0001) surface; (d) Adsorption energy of $NH_x$ molecules on both flat and step transition metals; (e) The volcano diagram of synthesis $NH_3$.[48] Reproduced with permission from ref. 48. Copyright 2012 RSC.

Nowadays, with the rapid development of computer technology, the first-principles calculation and volcano diagrams construction have been applied more and more on investigating electrocatalytic mechanisms and predicting promising catalysts of electrocatalytic NRR.[49, 50] Most of the studies combined theoretical calculation and experimental research in order to reveal the inner reaction and design high efficiency catalysts. Yu et al.[25] revealed the catalytic activities of different boron-doped graphene, that the $BC_3$ structure enabled the lowest energy barrier for $N_2$ electroreduction to $NH_3$. Tang et al.[51] reported a single transition metal atom sandwiched between hexagonal boron nitride (*h*-BN) and graphene sheets (BN/TM/G) which acts as an efficient single-atom catalyst for the electrochemical NRR based on first-principles calculations. A new reaction mechanism of NRR was proposed by Ling et al.,[52] based on the first-principles computation they proposed a surface hydrogenation mechanism for NRR in which can actuate NRR on catalysts with weaker $N_2$-binding strength at low potentials (Fig. 6). On another aspect, the promising catalysts predicting can effectively reduce workload during experimental researches, Zhang et al[53] investigated the catalytic performance of transition metal atom-anchored $C_2N$ monolayer electrocatalysts $TM_x@C_2N$ ($x$=1 or 2; TM=Ti, Mn, Fe, Co, Cu, Mo, Ru, Rh, Pd, Ag, Ir, Pt, or Au) for $N_2$ fixation. Chen et al.[54] evaluated catalytic performance of a series of single metal atoms supported on graphitic carbon nitride ($g$-$C_3N_4$) for NRR.

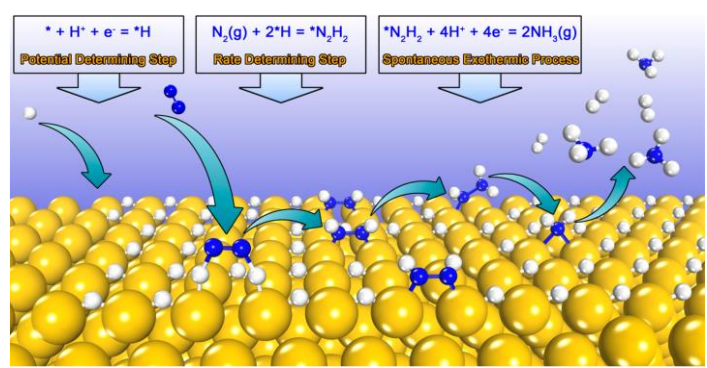

**Fig. 6** Schematic of surface-hydrogenation mechanism for NRR on noble-metal-based catalysts.[52] Reproduced with permission from ref. 52. Copyright 2019 ACS.

**3. Electrocatalytic process in ionic liquids**

**3.1 Characteristics of ILs**

Ionic liquids (ILs) are liquid state at or near room temperature and totally consist of anions and cations by ionic bonds. Compared to conventional solvents or electrolytes, ILs possess many advantages, such as low voltage, wide electrochemical window, large absorption capacity of some gases, non-volatile, non-flammable, high conductivity, high dissolubility and high stability.[55-59] Adopting hydrophobic, aprotic ILs or ILs and water mixtures as electrolytes can reduce available $H^+$ content and consequently suppress the competing reaction of HER.[35] In particular, the special feature of ILs is tunable chemical and physical properties by selecting appropriate cations and anions. Based on large amount of cations and anions, the number of ILs is theoretically estimated at about $10^{18}$.[60] Thus, the ILs could be designed to contain or not contain certain elements, such as N-containing or N-free, as well as be designed to improve hydrophobicity to suppress the HER process. Moreover, as a functional media, ILs can also be acted as an excellent solvent platform to design nanomaterials with unexpected morphologies and properties.[61, 62]

Due to the specific characteristics, ILs have been widely applied in many areas including electrochemistry, catalysis, biochemistry, organic synthesis, separation and so on. ILs have also attracted some attention in the field of electrocatalytic process and show with gratifying performance. We will introduce the application of ILs on the electrocatalytic process include $CO_2$ reduction reaction ($CO_2$RR), oxygen reduction reaction (ORR), *etc.* in this section, and the details of application and possible mechanism of ILs for electrocatalytic NRR process will be introduced in the next section.

**3.2 ILs as electrolytes and/or solvents**

In 2011, Rosen et al.[55] studied IL electrolyte for electrocatalytic process for the first time, and found that the IL could lower the free energy of intermediate as well as the overpotential. Then in 2014, Grills et al.[63] introduced ILs into the electrocatalytic $CO_2$RR process, they took ILs as both the solvent and electrolyte, and compared the reaction between pure IL of 1-ethyl-3-methylimidazolium tetracyanoborate ([Emim][TCB]) and acetonitrile ($CH_3CN$) for electrocatalytic $CO_2$RR with a homogeneous catalyst. It was found that when pure IL was used as the solvent and the electrolyte, the initial potential and overpotential were reduced by about 0.45 V (Fig. 7a), which significantly reduced the activation energy of the reaction. That was because the interactions of [Emim][TCB] and Re complex species promoted the chloride ligand dissociating faster, thereby reducing the activation energy and improving the electrocatalytic performance. Zhang's[64] group tested the activity of Mn–$C_3N_4$/CNT in both $CO_2$-saturated (1-butyl-3-methylimidazolium tetrafluoroborate ([Bmim]$BF_4$)/acetonitrile ($CH_3CN$)–$H_2O$) IL electrolyte and $CO_2$-saturated $KHCO_3$ electrolyte for electrocatalytic $CO_2$RR process, and obtained a wide overpotential range from 0.22 to 0.62 V with the CO FE of Mn–$C_3N_4$/CNT maintaining over 90%, as well as the highest $j_{CO}$ value was obtained at an overpotential of 0.62 V, which was higher than that of Mn–$C_3N_4$/CNT in the $KHCO_3$ electrolyte (as shown in Fig. 7b). Moreover, making full use of reactant is an effective way to improve reaction efficiency. For

CO2RR process, the raw material $CO_2$ must be sufficient to promote the reaction proceeding smoothly, and the target products can be continuously produced. In this study, the $CO_2$ solubility in [Bmim]$BF_4$ was much higher than that in $KHCO_3$ electrolyte. Meanwhile, the imidazolium cation could be arranged electrostatically at the cathode to favor $CO_2$ adsorption via [Bmim-$CO_2$]$^+$, and make the activation energy lower.[65] On the other side, the $CO_2$ reduction process also requires large activation energy and the competition of the HER also has great impacts on the $CO_2$ conversion efficiency. Several current studies have shown that adding ILs in electrolytes can not only increase the solubility of $CO_2$, but also effectively inhibit the HER competition.[66-68] Through DFT calculations, Feaster et al.[69] demonstrated that the [Emim]$^+$ cation could change the catalyst surface activity sites and suppress HER efficiently. As well as in electrocatalytic ORR process, the higher $O_2$ solubility could effectively improve the amount of reactant. Similarly, as shown by Wang et al.,[70] the higher solubility of $O_2$ could increase the utilization rate of active sites with transition metal-based porous ORR catalysts to enhance the electrocatalytic performance.

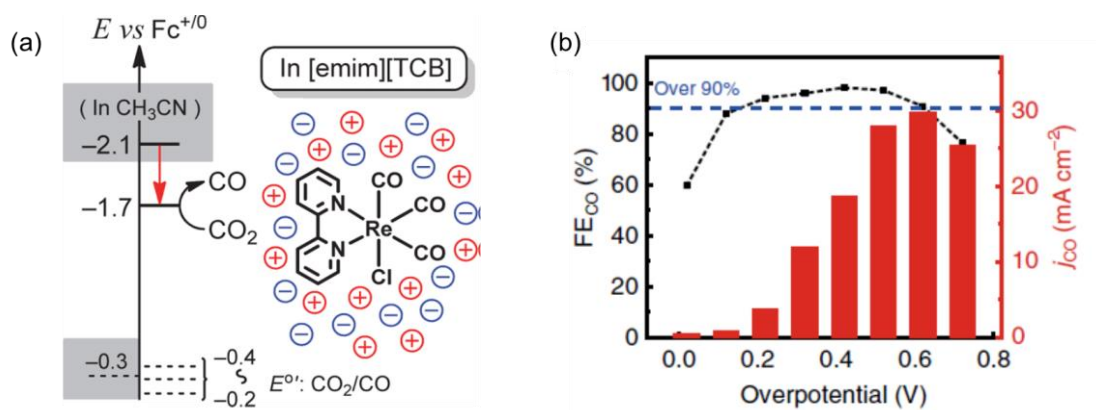

**Fig. 7** (a) Potential of IL as a solvent;[63] Reproduced with permission from ref. 63. Copyright 2014 ACS. (b) $FE_{CO}$ and $j_{CO}$ of Mn–$C_3N_4$/CNT at different potentials in the $CO_2$-saturated [Bmim]$BF_4$/$CH_3CN$-$H_2O$ electrolyte.[64] Reproduced with permission from ref. 64. Copyright 2020 Nature.

**3.3 ILs for catalysts synthesis**

With the unique nanostructures or microstructures and dynamical heterogeneities, ILs have been shown to tune the morphology and properties of the catalysts.[71, 72] ILs strategy is that take ILs as templates or precursors, and realize the fabrication of special constructions and morphologies. As early in 2008, Li et al.[73] took IL tetrabutylammonium hydroxide (TBAH) as a precursor to construct zinc oxide mesocrystals with the length of ca. 10 μm (as shown in Fig. 8a). Tang et al.[74] employed IL [Omim]Cl as a template and obtained Bohemithe (γ-AlOOH) with 3D flower-like architectures through a simple one-step synthesis process, and the morphology of flower-like γ-AlOOH is shown in Fig. 8b. Several studies have clarified that ILs could be used as the medium, template, precursor and reductant in the synthesis of nanomaterials with special structures, and such special structures may bring unexpected benefits to the electrocatalytic process.[75-78] For example, Kang et al.[79] took IL [Bmim][BF$_4$] as the template and synthesized the hierarchical mesoporous Prussian blue analogues (Cu-PBA), then applied in the CO$_2$RR process. The experimental results showed that there are more active sites existing in the mesoporous Cu-PBA and efficiently improve the activity and selectively for HCOOH.

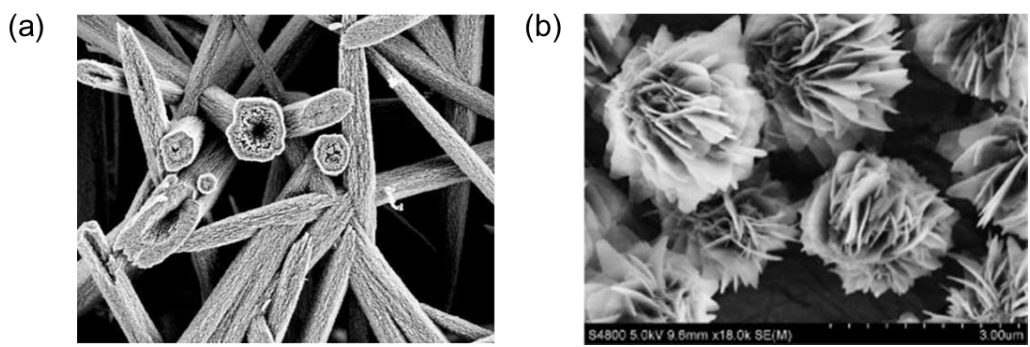

**Fig. 8** The special constructions and morphologies synthesized by ILs strategy: (a) zinc oxide mesocrystals;[73] Reproduced with permission from ref. 73. Copyright 2008 Wiley. (b) flower-like γ-AlOOH.[74] Reproduced with permission from ref. 74. Copyright 2013 Elsevier.

Moreover, ILs with the characteristics of hydrophobic and protic will play a crucial role in the electrocatalytic process. As early in 2010, Snyder et al.[80] put forward the material-synthesis strategies by adding ILs for the ORR process. They promoted an interesting catalysts synthesis strategy by adopting the secondary phase with higher O$_2$ solubility than external water phase to fill in the pores of 3D metal catalyst. With this structure, the O$_2$ will tend to stay in the pores no matter the O$_2$ solution gets equilibrium or not in external water phase. However, the selection of ILs was strict, and it must be highly O$_2$ soluble, hydrophobic, protic as well as with capillary forces, and finally the

IL of [MTBD][beti] reported by Luo et al.[81] was chosen.

The concept of solid catalyst with ionic liquid layer (SCILL) was presented by Kernchen et al. in 2007, they demonstrated that the solid ILs layer could improve the selectivity of heterogeneous catalysts effectively.[82] Similarly, the hydrophobic ILs could be also used as the surface modification material to coat on the catalyst and make the catalyst surface hydrophobic for the electrocatalytic ORR process, and the schematic diagram of the ILs coating formation process has been shown in Fig. 9. The ILs coatings was an uniform and thin ( < 2 nm) hydrophobic member to suppress the typical degradation reactions and prevent the destruction of catalyst surface[83, 84], as well as the hydrophobic of the ILs coatings created hydrophobic environment on catalyst surface that can repel the water molecules from products and electrolyte, protecting the active sites and suppress the formation of nonreactive oxygenated species.[84, 85]

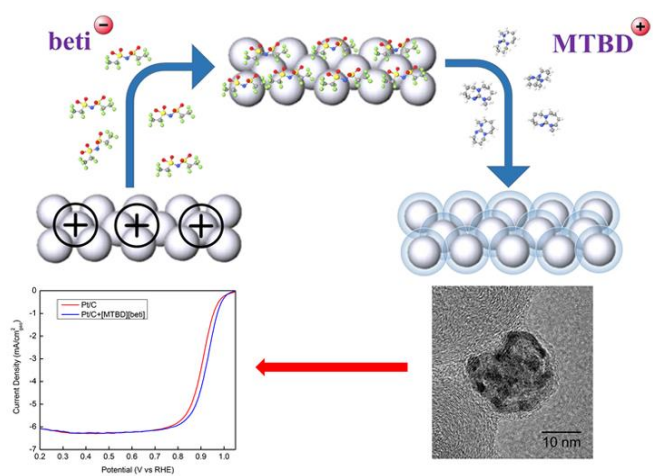

**Fig. 9** Schematic diagram of the sequential capacitive deposition (SCD) process.[83] Reproduced with permission from ref. 83. Copyright 2019 ACS.

Another effective catalyst synthesis strategy is to make higher specific surface area of catalyst by making the single-atom catalysts (SACs) dispersive. Ding et al.[86] reported an ionic liquid-stabilized single-atom catalysts (ILSSACs) strategy which was an electrical double layer of ILs supported single atom species and improved electrostatic stabilization to prevent the aggregation of isolated metal atoms, the schematic illustration of stabilization mechanism of SAC can be found in Fig. 10. With the addition of ILs, the specific surface area of metal could be enlarged efficiently and the catalytic performance was improved significantly.

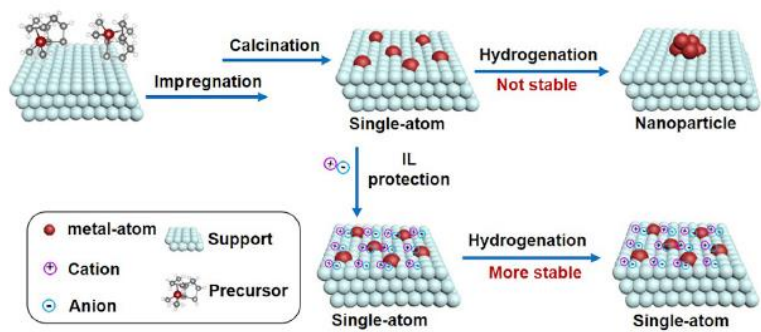

**Fig. 10** Schematic illustration of the preparation of SACs and the stabilization by ILs.[86] Reproduced with permission from ref. 86. Copyright 2019 Elsevier.

**4 ILs in electrocatalytic NRR process**

As described above, the ILs have been used in some fields of electrocatalytic process and enhanced the electrocatalytic performance effectively. For electrocatalytic NRR process, the first application of ILs was carried out in 2016 by Katayama et al.[87] They used hydrophobic IL 1-butyl-1-methylpyrrolidinium tris(pentafluoroethyl)trifluorophosphate ($[C_9H_{20}N][(C_2F_5)_3PF_3]$) as a support material and applied IL with a transition metal complex ($Cp_2TiCl_2$) as catalyst, then a solid polymer electrolysis (SPE) cell was prepared for the production of $NH_3$, and the schematic diagram is shown in Fig. 11a and b. This was the first time that hydrophobic ILs have been used in the design of SPE and NRR process under ambient conditions. The cyclic voltammogram of $Cp_2TiCl_2$ in $[C_9H_{20}N][(C_2F_5)_3PF_3]$ is shown in Fig. 11c. It can be seen that there are two reduction peaks in the negative potential region (-2.31 V and -2.54 V). The author conjectured that this was because the addition of the IL affected the redox of $Cp_2TiCl_2$. After evaluating the performance of electrochemical synthesis of $NH_3$, it was found that the yield of $NH_3$ can reach up to 27%, which is much higher than the previously reported yield of 1.45%. However, the deep mechanism and function of ILs were not clarified in the study. Hence, in this section we will focus on the possible role of ILs application in the NRR process.

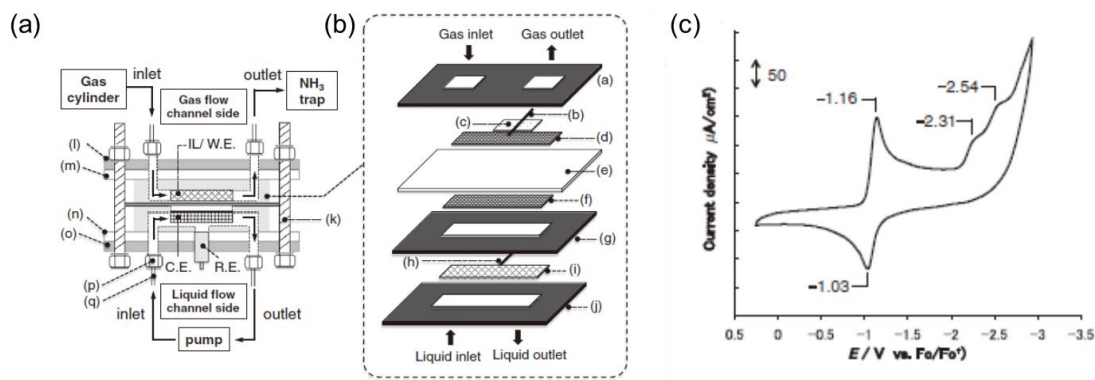

**Fig. 11** Scheme diagram of (a) SPE cell and (b) its interior structure; (c) Cyclic voltammograms of $Cp_2TiCl_2$ in $[C_9H_{20}N][(C_2F_5)_3PF_3]$.[87] Reproduced with permission from ref. 87. Copyright 2020 Elsevier.

**4.1 Absorption of $N_2$**

Solubility of $N_2$ is a very important factor for the formation of $NH_3$. Generally, electrolytes with high $N_2$ solubility tend to show high activation extents and low reduction barriers. At present, the most commonly used NRR solvent is $H_2O$, however, $N_2$ shows very little solubility in water because of its nonpolar, strong internally bonded and low polarizability nature. The extremely low solubility of $N_2$ in $H_2O$ which is only 0.66 mmol $L^{-1}$ resulting the poor NRR performance[62, 88]. Similar to $CO_2$, the solubility and diffusion coefficient of $N_2$ are very low in water, which will lead to significant changes of pH and $N_2$ concentration gradient and future affect the mass transfer.[89] Moreover, the aqueous solution often contains a large amount of $H^+$ proton sources, and it would increase the reaction effect of the competition of HER. Therefore, improving the solubility of $N_2$ in electrolyte is considered to be an effective way to increase the adsorption of $N_2$ at the active site, hence, to increase the performance of NRR process.

In recent years, the researches of $N_2$ solubility in ILs have been carried out. For example, Stevanovic and Gomes studied the solubility of various gases including $CO_2$, $N_2O$, $C_2H_6$ and $N_2$ in ILs 1-butyl-1-methylpyrrolidinium tris(pentafluoroethyl)trifluorophosphate $[C_1C_4Pyrro][eFAP]$ and trihexyl(tetradecyl)phosphonium tris(pentafluoroethyl)trifluorophosphate $[P_{66614}][eFAP]$. The authors found that $N_2$ was one order of magnitude less soluble than other gases in the two ILs, but was more soluble than in pure water.[90] To future obtained the mechanisms of dissolution of gases in fluorinated ILs, Gomes's group[91] studied the

ILs with partial fluorination on the cations, and found that fluorinated ILs exhibited higher $CO_2$ and $N_2$ mole fraction solubility. In comparison with the studies concerning $N_2$ solubility in low fluorinated ILs, it indicated that the higher fluorine content with ILs, the higher $N_2$ solubility.[92, 93] This provides possible evidence for ILs as a medium in NRR process. More importantly, the water content of ILs is lower than that of the aqueous solution, which will provide less $H^+$ source during the reaction, and effectively inhibit the generation of $H_2$.

Afterwards, inspired by previous studies, MacFarlane et al. firstly combined the $N_2$ solubility in ILs and the application of ILs in the electrochemical NRR. They synthesized a series of phosphonium-based ILs with highly fluorinated cations and anions[94] and 1-butyl-1-methylpyrrolidinium tris-(pentafluoroethyl) trifluorophosphate [$C_4$mpyr][eFAP]−fluorinated solvent mixtures,[95] and studied their related physicochemical properties including viscosity, ionic conductivity, ionicity, volumetric, and electrochemical stability. For fluorinated ILs, four cations and seven anions were synthesized, and they are listed in Fig. 12a. As shown in Fig. 12b, the relationship between perfluoroalkyl chain on the anion increasing and $N_2$ solubility was sufficiently relevant, and the $N_2$ solubility in all fluorinated ILs was improved significantly compared to water. For binary mixtures, the structures of [$C_4$mpyr][eFAP] and three fluorinated solvents are listed in Fig. 12c. Based on the experimental results, the authors demonstrated that the $N_2$ solubility increases with the increase of volume fraction of the fluorinated solvents, which is attributed to the weak interactions between these fluorinated solvents.[96] As seen from Fig. 12d, the $N_2$ solubility order is HFCP > FPEE > TFT, which is related to the cohesive energy density and fluorinated degree. Specifically speaking, HFCP has a high fluorinated degree and the lowest overall cohesive energy density, hence, it has a higher $N_2$ solubility. TFT contains only three fluorine atoms, thus, it shows a lower $N_2$ solubility. More importantly, the sufficient thermal stability and electrochemical stability of fluorinated ILs and [$C_4$mpyr][eFAP]−fluorinated solvent binary mixtures display potential application in the electrocatalytic NRR process.

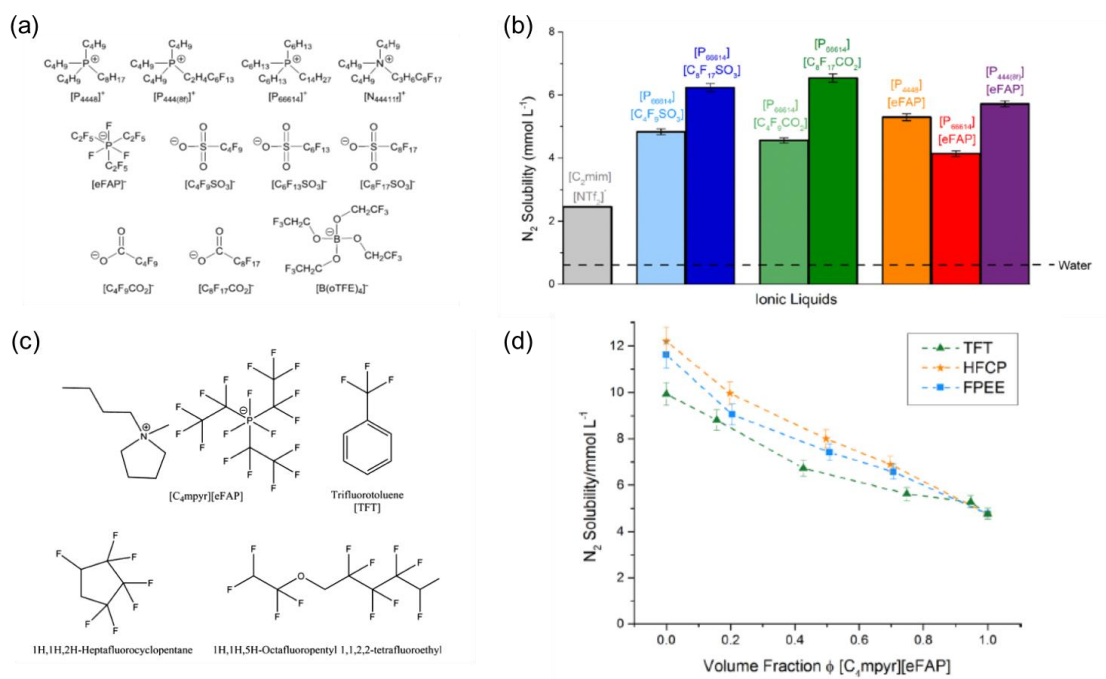

**Fig. 12** (a) The structures of ILs and (b) the $N_2$ solubilities in fluorinated ILs (mmol L$^{-1}$) at 30 °C and P=1 atm.[94] (c) Three fluorinated solvents and (d) the $N_2$ solubilities in [C$_4$mpyr][eFAP] binary mixtures at 30 °C and P=1 atm.[95] Reproduced with permission from ref. 94 and 95. Copyright 2018 and 2019 ACS.

### 4.2 Reduction of $N_2$

For electrocatalytic NRR process, the broken of nitrogen-nitrogen triple bond is almost impossible under ambient conditions, and the activation of $N_2$ is usually recognized as the rate-limiting step. Following early studies on the $N_2$ solubility in ILs, MacFarlane's group[35] compared the efficiencies of $NH_3$ production in several different ILs, and they found that FE was related to the $N_2$ solubility in ILs. For the ILs with the same anion, the FE of [P$_{66614}$][eFAP] was obtained up to 60% which was higher than that of [C$_4$mpyr][eFAP] (Fig. 13e). For comparison, they selected an IL with low $N_2$ solubility ([HMIM][NTf$_2$], of which the $N_2$ solubility is 0.017 mg g$^{-1}$), and found that the FE of $N_2$ reduction in this IL was much smaller, being 0.64%. Based on the DFT calculations on the interactions between $N_2$ and several anions ([PF$_6$]$^-$, Cl$^-$, [eFAF]$^-$, [BF$_4$]$^-$ and [NTf$_2$]$^-$), they found that the interaction between $N_2$ molecule and [eFAP]$^-$ was significantly higher than that with other ions, which was also one of the reasons for the higher solubility of $N_2$ (as shown in Fig. 13a-d).

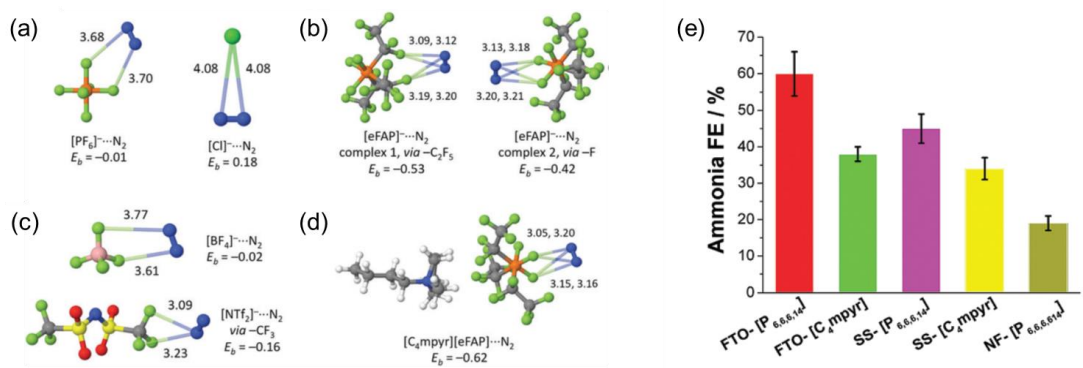

**Figure 13** (a–c) $N_2$ binding energies (in kcal mol$^{-1}$) and bond distances (in Å) for various anions; (d) The interaction between $N_2$ and [C$_4$mpyr][eFAP] after introducing the cation; (e) FE of $NH_3$ under different electrode and IL combinations.[35] Reproduced with permission from ref. 35. Copyright 2017 RSC.

Compared with ILs+H$_2$O electrolyte, with the same group of [eFAP]$^-$, they future designed an electrolyte with ILs+aprotic solvent ([C$_4$mpyr][eFAP]+1H,1H,5H-octafluoropentyl-1,1,2,2-tetrafluoroethyl ether (FPEE)). They found that employed aprotic ILs as electrolytes could make the supply of H$^+$ be regulated in favor of N$_2$ adsorption onto the catalytic sites. On the other hand, the catalyst with high specific surface area is conducive to improving the performance of catalysts, and N$_2$ will be activated by the coordination with the metal centers. The FE of NRR was as high as 32%, and the yield was 2.35×10$^{-11}$ mol s$^{-1}$ cm$^{-2}$. This also proved that under the action of the high gas solubility of ILs, the addition of aprotic solvents could better adjust H$^+$, which was beneficial to increase the selectivity of NRR, and greatly improve FE.[97]

On this basis, the interactions between bulk ILs and metal catalysts at the atomic level was simulated through classical molecular dynamics (MD) for the first time by Ortuño et al.[98] They compared the reaction energies of NRR (N$_2$*+H→N$_2$H*) and HER (H*+H→H$_2$*) on Ru surface with and without ILs, as shown in Fig. 14a. With IL, the NRR reaction energy was only 0.30 eV which was less than 0.64 eV without IL. Similarly, after adding IL, the HER reaction energy was 0.38 eV, while without IL, it was 0.52 eV. At the same time, it can be found that with IL the reaction energy of NRR is less than that of HER as shown in Fig. 14b. As a summary, the addition of IL improved both the activity and selectivity of NRR reaction.

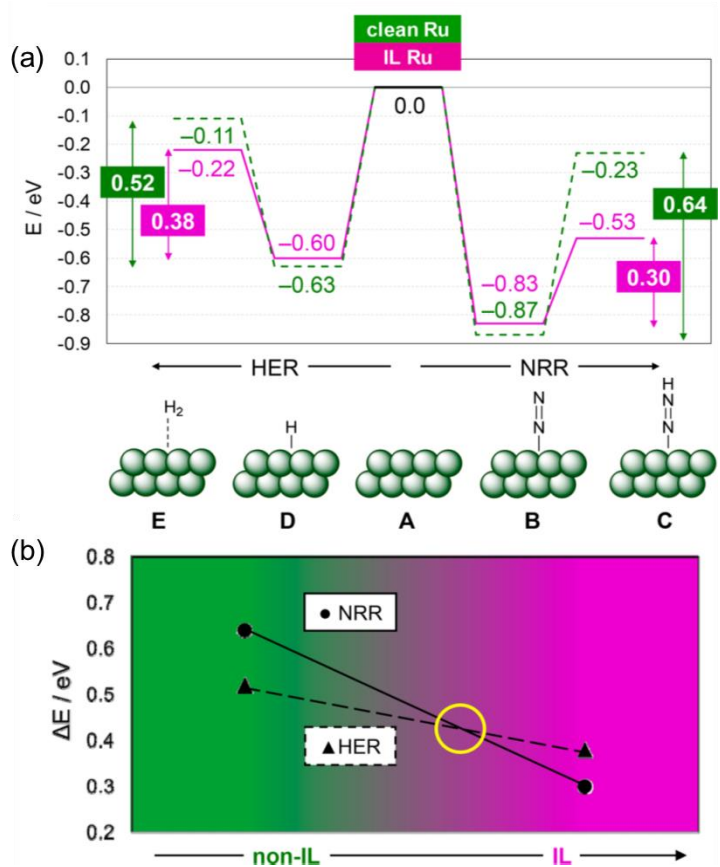

**Fig. 14** (a) Reaction energy for NRR and HER with and without IL; (b) Relative energy differences for NRR and HER with and without IL.[98] Reproduced with permission from ref. 98. Copyright 2019 ACS.

**4.3 Catalyst preparation in ILs**

As introduced in 3.2, the ILs strategy will bring unexpected advantages to electrocatalytic process. Recently, some studies have also been applied on electrocatalytic NRR process. For example, Li's group[99] proposed an ILs strategy for one-step synthesis *in situ* oxygen vacancy in $Fe_2O_3$ by taking IL n-octylammonium formate (OAF) as both the reaction medium and the structure-direction template. The results showed that the $Fe_2O_3$-IL has a smaller size, a larger surface area, and more active sites for NRR (as shown in Fig. 15). Moreover, the rich oxygen vacancy in $Fe_2O_3$-IL could generate lone pair electrons on the catalyst surface to induce the polarity of $N_2$ and favor the adsorption and activation of $N_2$. In 2020, the authors once again adopted ILs strategy to synthesize 3D Rh particles with nanowires subunits which possess more active sites and provide more transfers channels for ions or electrons (as shown in Fig. 16). In the study, they also selected the IL OAF as the solvent, reductant and template, and the formic anion of IL firstly reduced $Rh^{3+}$ and then the template function of IL fabricated the final 3D Rh nanostructure with nanowires subunits.[36] Note that, the performances of the electrocatalytic NRR were improved effectively in both studies.

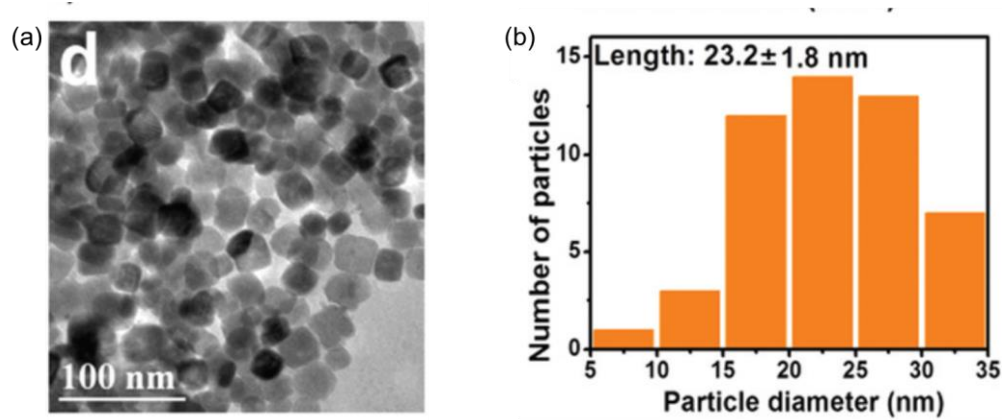

**Fig. 15** (a) TEM and (b) particle size distribution of $Fe_2O_3$-IL.[99] Reproduced with permission from ref. 99. Copyright 2019 RSC.

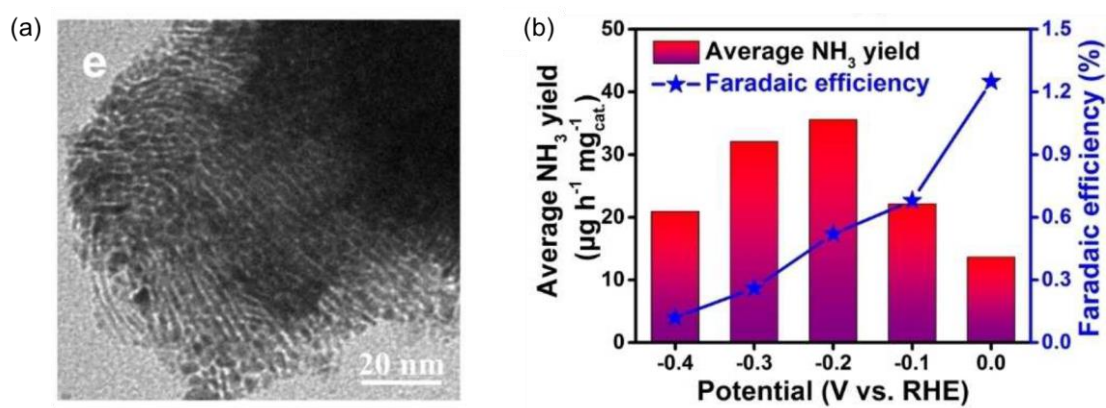

**Fig. 16** (a) TEM of 3D Rh and (b) average $NH_3$ yield and FE of 3D Rh catalyst.[36] Reproduced with permission from ref. 36. Copyright 2020 ACES.

## 5. Conclusions and perspectives

Electrocatalytic NRR is a promising alternative synthesis method of the traditional Haber–Bosch process under mild conditions. Moreover, the ILs are widely applied in many areas due to their unique and excellent characteristics. However, the combination studies of electrocatalytic NRR process and ILs are somehow limited. Although electrocatalytic NRR is a shining star in academic world, the industrial application is still a long way to go. To some extent, the introduction of ILs has been demonstrated to be viable and to improve the $NH_3$ yield and FE. Many studies have also clarified the effective role of ILs in promoting the process of $CO_2RR$ and ORR with both good experimental and theoretical results. Here, we summarize the possible function and mechanism of ILs in the electrocatalytic NRR process as follows: (1) **Higher $N_2$ solubility.** The $N_2$ solubility in some kinds of ILs is higher than that in water. Choosing ILs or ILs mixtures as electrolytes is able to improve the $N_2$ content in solvent and potentially improve the corresponding $NH_3$ yield. (2) **Inhibiting competition reaction.** Choosing aprotic and hydrophobic ILs as electrolytes in whole or in part is capable of reducing the water amount as well as reducing proton contents to suppress the competing of HER. (3) **Better $N_2$ adsorption.** On the one hand, ILs could make the catalyst metal more dispersive instead of aggregation, resulting in larger specific surface areas, and more $N_2$ adsorption sites. On the other hand, aprotic ILs could regulate the supplement of $H^+$ to favor $N_2$ adsorption onto the catalytic sites. (4) **Construction special morphologies.** Appropriate ILs are selected as the solvent, medium, reductant or template to synthesize catalyst with special constructions and morphologies. The special construction of catalysts could show special catalytic properties. (5) **Non-volatility and stability.** The sufficient thermal stability and electrochemical stability of ILs will make them having potential applications on electrocatalytic NRR process. Moreover, the characteristic of non-volatility will keep the system stable and make the products more pure. In short, the following types of ILs should be considered in the vein of catalysis: (1) with low viscosity should be chosen, (2) aprotic and hydrophobic ILs should be chosen, and (3) the ILs should have little or without N atom, and with high $N_2$ solubility.

The complex structure of ILs makes it difficult to conduct intensive and unambiguous studies in the NRR process, as well as large-scale application. In the future, the mechanism and internal process research of NRR and ILs are important, which need more in-deep studies. On the one hand, with the rapid development of computer technology, theoretical calculations will be fully utilized to accelerate the progress of NRR. Calculations of reaction pathways and adsorption energies are far from enough, and the much deeper research including catalysts surface construction, electrode design, intermediates identification and the interactions between electrolytes and electrodes *etc.* should be carried out by combining theoretical calculations and experimental investigations. On the other hand, modern *in situ* and *operando* technology will play a critical role in the development of NRR. The *in situ* and *operando* technologies could realize *in situ*, real-time, and zero-decrement tests, and the surface changes, intermediates reactions, phase changes, *etc.* all could be directly observed. In the follow-up studies, we believe that the mechanism and internal process of NRR no matter with or without the ILs can be a better understanding under the advanced theoretical calculations and *in situ*/*operando* technologies. Meanwhile, the detailed function and mechanism of ILs in the electrocatalytic NRR will be revealed, and the electrocatalytic NRR with more efficient and higher selectivity will be around the corner.

# References

1. X. Cui, C. Tang and Q. Zhang, *Adv. Energy Mater.*, 2018, **8**, 1800369.
2. R. Schlögl, *Angew. Chem. Int. Ed.*, 2003, **42**, 2004-2008.
3. A. Klerke, C. H. Christensen, J. K. Nørskov and T. Vegge, *J. Mater. Chem.*, 2008, **18**, 2304-2310.
4. C. J. Van der Ham, M. T. Koper and D. G. Hetterscheid, *Chem. Soc. Rev.*, 2014, **43**, 5183-5191.
5. J. G. Chen, R. M. Crooks, L. C. Seefeldt, K. L. Bren, R. M. Bullock, M. Y. Darensbourg, P. L. Holland, B. Hoffman, M. J. Janik and A. K. Jones, *Science*, 2018, **360**, eaar6611.
6. R. F. Service, *Science*, 2018, aau7489.
7. R. F. Service, *Science,* 2018, **361**, 120-123.
8. R. M. Martínez-Espinosa, J. A. Cole, D. J. Richardson and N. J. Watmough, *Biochem. Soc. Trans.*, 2011, **39**, 175-178.
9. B. K. Burgess and D. J. Lowe, *Chem. Rev.*, 1996, **96**, 2983-3012.
10. T. Kandemir, M. E. Schuster, A. Senyshyn, M. Behrens and R. Schlögl, *Angew. Chem. Int. Ed.*, 2013, **52**, 12723-12726.
11. H. Liu, *Chinese J. Catal.*, 2014, **35**, 1619-1640.
12. V. Smil, *Sci. Am.*, 1997, **277**, 76-81.
13. B. Cui, J. Zhang, S. Liu, X. Liu, W. Xiang, L. Liu, H. Xin, M. J. Lefler and S. Licht, *Green Chem.*, 2017, **19**, 298-304.
14. K. Arashiba, E. Kinoshita, S. Kuriyama, A. Eizawa, K. Nakajima, H. Tanaka, K. Yoshizawa and Y. Nishibayashi, *J. Am. Chem. Soc.*, 2015, **137**, 5666-5669.
15. Z. Wei, Y. Zhang, S. Wang, C. Wang and J. Ma, *J. Mater. Chem. A*, 2018, **6**, 13790-13796.
16. H. Dong, G. Zeng, L. Tang, C. Fan, C. Zhang, X. He and Y. He, *Water Res.*, 2015, **79**, 128-146.
17. J. C. Colmenares, R. Luque, J. M. Campelo, F. Colmenares, Z. Karpiński and A. A. Romero, *Materials*, 2009, **2**, 2228-2258.
18. G.-F. Chen, Y. Yuan, H. Jiang, S.-Y. Ren, L.-X. Ding, L. Ma, T. Wu, J. Lu and H. Wang, *Nat. Energy*, 2020, **5**, 605-613.
19. Y.-C. Hao, Y. Guo, L.-W. Chen, M. Shu, X.-Y. Wang, T.-A. Bu, W.-Y. Gao, N. Zhang, X. Su and X. Feng, *Nat. Catal.*, 2019, **2**, 448-456.
20. E. E. Van Tamelen and D. A. Seeley, *J. Am. Chem. Soc.*, 1969, **91**, 5194-5194.
21. S. F. McWilliams and P. L. Holland, *Acc. Chem. Res.*, 2015, **48**, 2059-2065.
22. J. Qian, Q. An, A. Fortunelli, R. J. Nielsen and W. A. Goddard III, *J. Am. Chem. Soc.*, 2018, **140**, 6288-

6297.

23. B. Yang, W. Ding, H. Zhang and S. Zhang, *Energ. Environ. Sci.*, 2020, **14**, 672-687.
24. J. Deng, J. A. Iñiguez and C. Liu, *Joule*, 2018, **2**, 846-856.
25. X. Yu, P. Han, Z. Wei, L. Huang, Z. Gu, S. Peng, J. Ma and G. Zheng, *Joule*, 2018, **2**, 1610-1622.
26. X. Wang, W. Wang, M. Qiao, G. Wu, W. Chen, T. Yuan, Q. Xu, M. Chen, Y. Zhang and X. Wang, *Sci. Bull.*, 2018, **63**, 1246-1253.
27. L. Li, C. Tang, B. Xia, H. Jin, Y. Zheng and S.-Z. Qiao, *ACS Catal.*, 2019, **9**, 2902-2908.
28. L. Tan, N. Yang, X. Huang, L. Peng, C. Tong, M. Deng, X. Tang, L. Li, Q. Liao and Z. Wei, *Chem. Commun.*, 2019, **55**, 14482-14485.
29. L. Guangkai, J. Haeseong, L. Zijian, W. Jia, J. Xuqiang, K. Min Gyu, L. Xien and C. Jaephil, *Green Energy Environ.*, 2020, DOI: https://doi.org/10.1016/j.gee.2020.11.004.
30. J. Zhang, X. Tian, M. Liu, H. Guo, J. Zhou, Q. Fang, Z. Liu, Q. Wu and J. Lou, *J. Am. Chem. Soc.*, 2019, **141**, 19269-19275.
31. Y. Wang, M. m. Shi, D. Bao, F. l. Meng, Q. Zhang, Y. t. Zhou, K. h. Liu, Y. Zhang, J. z. Wang and Z. w. Chen, *Angew. Chem. Int. Ed.*, 2019, **58**, 9464-9469.
32. Y. Yang, S. Q. Wang, H. Wen, T. Ye, J. Chen, C. P. Li and M. Du, *Angew. Chem. Int. Ed.*, 2019, **131**, 15506-15510.
33. B. H. Suryanto, D. Wang, L. M. Azofra, M. Harb, L. Cavallo, R. Jalili, D. R. Mitchell, M. Chatti and D. R. MacFarlane, *ACS Energy Lett.*, 2018, **4**, 430-435.
34. Q. Zhang, B. Liu, L. Yu, Y. Bei and B. Tang, *ChemCatChem*, 2020, **12**, 334-341.
35. F. Zhou, L. M. Azofra, M. Ali, M. Kar, A. N. Simonov, C. McDonnell-Worth, C. Sun, X. Zhang and D. R. MacFarlane, *Energ. Environ. Sci.*, 2017, **10**, 2516-2520.
36. T. Chen, S. Liu, H. Ying, Z. Li and J. Hao, *Chem.–Asian J.*, 2020, **15**, 1081-1087.
37. W. H. Guo, K. X. Zhang, Z. B. Liang, R. Q. Zou and Q. Xu, *Chem. Soc. Rev.*, 2019, **48**, 5658-5716.
38. B. H. R. Suryanto, H. L. Du, D. B. Wang, J. Chen, A. N. Simonov and D. R. MacFarlane, *Nat. Catal.*, 2019, **2**, 290-296.
39. G. Qing, R. Ghazfar, S. T. Jackowski, F. Habibzadeh, M. M. Ashtiani, C. P. Chen, M. R. Smith and T. W. Hamann, *Chem. Rev.*, 2020, **120**, 5437-5516.
40. C. Tang and S. Z. Qiao, *Chem. Soc. Rev.*, 2019, **48**, 3166-3180.
41. H. Cheng, P. X. Cui, F. R. Wang, L. X. Ding and H. H. Wang, *Angew. Chem. Int. Ed.*, 2019, **58**, 15541-15547.
42. J. H. Montoya, C. Tsai, A. Vojvodic and J. K. Nørskov, *ChemSusChem*, 2015, **8**, 2180-2186.
43. G. Ertl, *Angew. Chem. Int. Ed.*, 1990, **29**, 1219-1227.
44. G. Ertl, in *Catalysis*, Springer, 1983, pp. 209-282.
45. X.-F. Li, Q.-K. Li, J. Cheng, L. Liu, Q. Yan, Y. Wu, X.-H. Zhang, Z.-Y. Wang, Q. Qiu and Y. Luo, *J. Am. Chem. Soc.*, 2016, **138**, 8706-8709.
46. Y. Abghoui and E. Skúlason, *Catal. Today*, 2017, **286**, 78-84.
47. X. Liao, R. Lu, L. Xia, Q. Liu, H. Wang, K. Zhao, Z. Wang and Y. Zhao, *Energy Environ. Mater.*, 2021, DOI: https://doi.org/10.1002/eem2.12204.
48. E. Skulason, T. Bligaard, S. Gudmundsdóttir, F. Studt, J. Rossmeisl, F. Abild-Pedersen, T. Vegge, H. Jónsson and J. K. Nørskov, *PCCP*, 2012, **14**, 1235-1245.
49. J. K. Nørskov, T. Bligaard, J. Rossmeisl and C. H. Christensen, *Nat. Chem.*, 2009, **1**, 37-46.
50. Z. Seh, J. Kibsgaard, C. Dickens, I. Chorkendorff, J. Norskov and T. F. Jaramillo, *Science*, 2017, **355**, eaad4998.

51. S. Tang, Q. Dang, T. Liu, S. Zhang, Z. Zhou, X. Li, X. Wang, E. Sharman, Y. Luo and J. Jiang, *J. Am. Chem. Soc.*, 2020, **142**, 19308-19315.
52. C. Ling, Y. Zhang, Q. Li, X. Bai, L. Shi and J. Wang, *J. Am. Chem. Soc.*, 2019, **141**, 18264-18270.
53. X. Zhang, A. Chen, Z. Zhang and Z. Zhou, *J. Mater. Chem. A*, 2018, **6**, 18599-18604.
54. Z. Chen, J. Zhao, C. R. Cabrera and Z. Chen, *Small Methods*, 2019, **3**, 1800368.
55. B. A. Rosen, A. Salehi-Khojin, M. R. Thorson, W. Zhu, D. T. Whipple, P. J. Kenis and R. I. Masel, *Science*, 2011, **334**, 643-644.
56. M. Alvarez-Guerra, J. Albo, E. Alvarez-Guerra and A. Irabien, *Energ. Environ. Sci.*, 2015, **8**, 2574-2599.
57. M. Asadi, K. Kim, C. Liu, A. V. Addepalli, P. Abbasi, P. Yasaei, P. Phillips, A. Behranginia, J. M. Cerrato, R. Haasch, P. Zapol, B. Kumar, R. F. Klie, J. Abiade, L. A. Curtiss and A. Salehi-Khojin, *Science*, 2016, **353**, 467-470.
58. P. Hapiot and C. Lagrost, *Chem. Rev.*, 2008, **108**, 2238-2264.
59. J. Li, F. Li, L. Zhang, H. Zhang, L. Ulla and X. Ji, *Green Chem. Eng.*, 2021, **2**, 253-265.
60. Y. Chen and T. Mu, *Green Chem. Eng.*, 2021, **2**, 174-186.
61. K. S. Egorova, E. G. Gordeev and V. P. Ananikov, *Chem. Rev.*, 2017, **117**, 7132-7189.
62. C. Zhang, B. Xin, Z. Xi, B. Zhang, Z. Li, H. Zhang, Z. Li and J. Hao, *ACS Sustain. Chem. Eng.*, 2018, **6**, 1468-1477.
63. D. C. Grills, Y. Matsubara, Y. Kuwahara, S. R. Golisz, D. A. Kurtz and B. A. Mello, *J. Phys. Chem. Lett.*, 2014, **5**, 2033-2038.
64. J. Feng, H. Gao, L. Zheng, Z. Chen, S. Zeng, C. Jiang, H. Dong, L. Liu, S. Zhang and X. Zhang, *Nat. Commun.*, 2020, **11**, 1-8.
65. B. A. Rosen, J. L. Haan, P. Mukherjee, B. Braunschweig, W. Zhu, A. Salehi-Khojin, D. D. Dlott and R. I. Masel, *J. Phys. Chem. C*, 2012, **116**, 15307-15312.
66. W. Lu, B. Jia, B. Cui, Y. Zhang, K. Yao, Y. Zhao and J. Wang, *Angew. Chem. Int. Ed.*, 2017, **56**, 11851-11854.
67. J. Feng, S. Zeng, H. Liu, J. Feng, H. Gao, L. Bai, H. Dong, S. Zhang and X. Zhang, *ChemSusChem*, 2018, **11**, 3191-3197.
68. F. Li, F. Mocci, X. Zhang, X. Ji and A. Laaksonen, *Chin. J. Chem. Eng.*, 2020, **31**, 75-93.
69. J. T. Feaster, A. L. Jongerius, X. Liu, M. Urushihara, S. A. Nitopi, C. Hahn, K. Chan, J. K. Nørskov and T. F. Jaramillo, *Langmuir*, 2017, **33**, 9464-9471.
70. M. Wang, H. Zhang, G. Thirunavukkarasu, I. Salam, J. R. Varcoe, P. Mardle, X. Li, S. Mu and S. Du, *ACS Energy Lett.*, 2019, **4**, 2104-2110.
71. Y.-L. Wang, B. Li, S. Sarman, F. Mocci, Z.-Y. Lu, J. Yuan, A. Laaksonen and M. D. Fayer, *Chem. Rev.*, 2020, **120**, 5798-5877.
72. X. Kang, X. Sun, X. Ma, P. Zhang, Z. Zhang, Q. Meng and B. Han, *Angew. Chem. Int. Ed.*, 2017, **56**, 12683-12686.
73. Z. Li, A. Geßner, J. P. Richters, J. Kalden, T. Voss, C. Kübel and A. Taubert, *Adv. Mater.*, 2008, **20**, 1279-1285.
74. Z. Tang, J. Liang, X. Li, J. Li, H. Guo, Y. Liu and C. Liu, *J. Solid State Chem.*, 2013, **202**, 305-314.
75. X. Duan, J. Ma, J. Lian and W. Zheng, *CrystEngComm*, 2014, **16**, 2550-2559.
76. X. Kang, X. Sun and B. Han, *Adv. Mater.*, 2016, **28**, 1011-1030.
77. X.-D. Liu, H. Chen, S.-S. Liu, L.-Q. Ye and Y.-P. Li, *Mater. Res. Bull.*, 2015, **62**, 217-221.
78. B. Zhang, Y. Xue, A. Jiang, Z. Xue, Z. Li and J. Hao, *ACS Appl. Mater. Inter.*, 2017, **9**, 7217-7223.
79. X. Kang, X. Sun, Q. Zhu, X. Ma, H. Liu, J. Ma, Q. Qian and B. Han, *Green Chem.*, 2016, **18**, 1869-1873.

80. J. Snyder, T. Fujita, M. Chen and J. Erlebacher, *Nat. Mater.*, 2010, **9**, 904-907.
81. H. Luo, G. A. Baker, J. S. Lee, R. M. Pagni and S. Dai, *J. Phys. Chem. B*, 2009, **113**, 4181-4183.
82. U. Kernchen, B. Etzold, W. Korth and A. Jess, *Chem. Eng. Technol.*, 2007, **30**, 985-994.
83. Y. Li, J. Hart, L. Profitt, S. Intikhab, S. Chatterjee, M. Taheri and J. Snyder, *ACS Catal.*, 2019, **9**, 9311-9316.
84. M. George, G.-R. Zhang, N. Schmitt, K. Brunnengraber, D. J. Sandbeck, K. J. Mayrhofer, S. Cherevko and B. J. J. A. c. Etzold, *ACS Catal.*, 2019, **9**, 8682-8692.
85. G. R. Zhang, M. Munoz and B. J. Etzold, *Angew. Chem. Int. Ed.*, 2016, **55**, 2257-2261.
86. S. Ding, Y. Guo, M. J. Hülsey, B. Zhang, H. Asakura, L. Liu, Y. Han, M. Gao, J.-y. Hasegawa and B. Qiao, *Chem*, 2019, **5**, 3207-3219.
87. A. Katayama, T. Inomata, T. Ozawa and H. Masuda, *Electrochem. Commun.*, 2016, **67**, 6-10.
88. R. Battino, T. R. Rettich and T. Tominaga, *J. Phys. Chem. Ref. Data*, 1984, **13**, 563-600.
89. E. L. Clark, J. Resasco, A. Landers, J. Lin, L.-T. Chung, A. Walton, C. Hahn, T. F. Jaramillo and A. T. Bell, *ACS Catal.*, 2018, **8**, 6560-6570.
90. S. Stevanovic and M. C. Gomes, *J. Chem. Thermodyn.*, 2013, **59**, 65-71.
91. D. Almantariotis, A. Pensado, H. Gunaratne, C. Hardacre, A. Pádua, J.-Y. Coxam and M. C. Gomes, *J. Phys. Chem. B*, 2017, **121**, 426-436.
92. J. L. Anderson, J. K. Dixon and J. F. Brennecke, *Acc. Chem. Res.*, 2007, **40**, 1208-1216.
93. J. Blath, M. Christ, N. Deubler, T. Hirth and T. Schiestel, *Chem. Eng. J.*, 2011, **172**, 167-176.
94. C. S. Kang, X. Zhang and D. R. MacFarlane, *J. Phys. Chem. C*, 2018, **122**, 24550-24558.
95. C. S. Kang, X. Zhang and D. R. MacFarlane, *J. Phys. Chem. C*, 2019, **123**, 21376-21385.
96. J. D. Dunitz, *ChemBioChem*, 2004, **5**, 614-621.
97. B. H. R. Suryanto, C. S. M. Kang, D. Wang, C. Xiao, F. Zhou, L. M. Azofra, L. Cavallo, X. Zhang and D. R. MacFarlane, *ACS Energy Lett.*, 2018, **3**, 1219-1224.
98. M. A. Ortuno, O. Holloczki, B. Kirchner and N. Lopez, *J. Phys. Chem. Lett.*, 2019, **10**, 513-517.
99. C. Zhang, S. Liu, T. Chen, Z. Li and J. Hao, *Chem. Commun.*, 2019, **55**, 7370-7373.